\documentclass{aa}
\usepackage{graphics}
\usepackage{psfig}
\begin{document}

%   \thesaurus{11     % A&A Section 11: Galaxies
%              (11.09.1 Mrk 54;  % Galaxies: individual: Mrk 54
%               11.09.3;  % intergalactic medium,
%               11.09.4;  % Galaxies: ISM,
%               11.19.3;  % Galaxies: starburst,
%               12.04.2;  % diffuse radiation,
%               13.21.1)} % Ultraviolet: galaxies.
%
   \title{  Constraints on the Lyman continuum radiation from galaxies:
  first results with FUSE on Mrk 54
 \thanks{Based on observations 
 made with the NASA-CNES-CSA Far Ultraviolet 
 Spectroscopic Explorer. FUSE is operated for NASA by the Johns Hopkins 
 University under NASA contract NAS5-32985.}}

   \subtitle{}

   \author{J.-M. Deharveng
          \inst{1}
          \and
          V. Buat\inst{1}
           \and
          V. Le Brun\inst{1}
            \and
          B. Milliard\inst{1}
             \and
          D. Kunth\inst{2}      
              \and
          J. M. Shull\inst{3}
               \and
           C. Gry\inst{1,4}
             }

   \offprints{J.-M. Deharveng, \email{jean-michel.deharveng@astrsp-mrs.fr} }

   \institute{Laboratoire d'Astrophysique de Marseille, 
              Traverse du Siphon, Les Trois Lucs,
              BP 8, 13376 Marseille Cedex 12, France
               \and
             Institut d'Astrophysique de Paris,
             98 bis Boulevard Arago, 75014 Paris, France
               \and            
             Center for Astrophysics and Space Astronomy, Department
             of Astrophysical and Planetary Sciences, University of Colorado,
             Boulder, CO 80309, USA
             \and 
             ISO Data Center, ESA Astrophysics Division, P.O. Box 50727,
             28080 Madrid, Spain
                    }

   \date{Received  .... 2001 / Accepted .... 2001}

   \titlerunning{Lyman continuum radiation from galaxies}
   \authorrunning{Deharveng et al.}

   \abstract{We present {\it Far Ultraviolet Spectroscopic Explorer}
   observations of the star-forming galaxy Mrk 54 at z = 0.0448. The 
   Lyman continuum radiation is not detected above the
   \ion{H}{i} absorption edge 
   in our Galaxy. An upper limit is evaluated by comparison with 
   the background
   measured in regions of the detector adjacent to the observed 
   spectrum. A spectral window of 16 \AA,
   reasonably free of additional \ion{H}{i} Lyman series line absorption
   is used. No correction is needed for molecular hydrogen 
   absorption in our Galaxy but a foreground extinction of 0.29 mag is 
   accounted for. An upper limit of $6.15 \times 10^{-16}$ 
   erg cm$^{-2}$ s$^{-1}$ A$^{-1}$ is obtained for the flux 
   at $\approx$ 900 \AA~ in the rest frame of Mrk 54. By comparison with 
   the number of ionizing photons derived from the H$\alpha$
   flux, this limit translates into an upper limit of 
   $f_{\mathrm{esc}}< 0.062$ for the fraction of Lyman continuum 
   photons that escape the galaxy without being absorbed by 
   interstellar material. This limit compares with the limits 
   obtained in three other nearby galaxies and is 
   compatible with the escape fractions predicted by models.
   The upper limits obtained in nearby galaxies contrasts with the detection 
   of Lyman continuum flux in the composite spectrum of Lyman-break
   galaxies at z $\approx$ 3.4. The difficulties and implications
   of a comparison are discussed.
   }

   \maketitle

%
%________________________________________________________________

\section{Introduction}

     It is not yet clear whether hot and massive stars forming 
   in galaxies contribute 
  significantly to the ionizing background radiation in the universe,
  and how this contribution evolves as a function of redshift
  (e.g. Bechtold et al. \cite{bec}, Miralda-Escud\'e \& Ostriker
   \cite{mir}, Meiksin \& Madau \cite{mei}, 
   Madau \& Shull \cite{mad1}, Haardt \& Madau \cite{haa}, 
   Shull et al. \cite{shu1}). Specifically, 
   at redshifts z $>$ 3, the early formation of galaxies 
  is expected to compensate for the decline in the quasar contribution
  and to play a role in the re-ionization of the IGM (e.g. Madau et al. 
   \cite{mad2}).

     Direct observations of galaxies below the Lyman break have been scarce
  so far. With the Hopkins Ultraviolet Telescope 
  (HUT), Leitherer et al. (\cite{lei1}), Hurwitz et al. (\cite{hur}) 
  obtained upper limits on the Lyman continuum (LyC) radiation 
  in four 
  nearby star-forming galaxies. By comparison  
  with the H$\alpha$
  emission these data were  
  interpreted in terms of limits on the LyC escape fraction, a parameter 
  giving the fraction of hydrogen-ionizing photons effectively 
  released into the IGM.
  Recently Steidel et al. (\cite{ste}) reported the detection of the 
  LyC radiation in a composite spectrum of Lyman break galaxies
   at z $\sim$ 3.4 that was
  also interpreted in terms of LyC escape fraction but by comparison 
  with the (1500 \AA) UV continuum. A preliminary account of 
  the Space Telescope Imaging Spectrograph
  (STIS) observations of galaxies in the Hubble Deep Field (HDF),
  with implications 
  on the LyC escape fraction, has been presented by Ferguson (\cite{fer})
  while this paper was nearing completion. 

  In the absence of a large number of observations
  that would directly provide the LyC luminosity function and 
  the  LyC luminosity density of galaxies, the LyC escape fraction
  is seen as a crucial parameter.
  Combined with the H$\alpha$ or UV luminosity densities of galaxies
  or with 
  the stellar ionizing radiation calculated from evolutionary 
  synthesis models (e.g. Bruzual 
  \& Charlot \cite{bru}, Leitherer et al. \cite{lei2}) or  
  from the rate of chemical enrichment in the universe (Cowie \cite{cow},
  Songaila et al. \cite{son}, Madau \& Shull \cite{mad1}), 
  the LyC escape fraction provides 
  the amount of ionizing radiation
  effectively released into the IGM by the galaxies.
  As this parameter encapsulates a number 
  of complex and random factors it is probably highly variable from 
  galaxy to galaxy; its full understanding and sound utilisation   
  would also require a large number of observations.  

     The {\it Far Ultraviolet Spectroscopic Explorer (FUSE)} 
  has recently opened again an access into the far ultraviolet 
  down to 905 \AA, allowing the possibility to observe
  the LyC radiation of low redshift galaxies above the 
  912 \AA~ Lyman limit of \ion{H}{i}
   photoelectric absorption in our Galaxy. 
  As shown by the re-analysis of 
  HUT data by Hurwitz et al. (\cite{hur}), getting rid of
  photoelectric absorption in our Galaxy above the Lyman edge is not enough
  and gas-phase
  absorption in our Galaxy, essentially from convergent \ion{H}{i} 
  Lyman series 
  and the Lyman and Werner bands of 
  molecular hydrogen has to be accounted for
  in addition to dust extinction. 
  It is an area where the high spectral resolution of 
  {\it FUSE} can help even though the increased background contribution per 
  wavelength unit makes the detection of faint continua  
  more difficult than at low resolution.   
  In the following we report the {\it FUSE} observations of a star-forming
  galaxy Mrk 54, at a redshift z = 0.0448 that places the Lyman break
  at 952.5 \AA~ 
  above the Galactic Ly$\delta$ absorption feature, reasonably beyond 
  the convergent \ion{H}{i} Lyman series.     

\section{Observations and Data Processing}

   The observations of Mrk 54 (GI program 
  A052) were obtained on Feb 19 -
  20, 2000 in the time-tagged photon address mode with
 the object in the large (30\arcsec $\times$ 30\arcsec)
 aperture. The observations were split into 15 exposures
 between occultation periods and passages through 
 the South Atlantic Anomaly, and generally extended 
 over spacecraft night and day.
 The total duration was 27502 seconds.
 Details of the {\it FUSE} instrument and on-orbit performance
 have been given by Moos et al. (\cite{moo}) and Sahnow et al. (\cite{sah}).

 The summed and calibrated spectrum of Mrk 54 
 is consistent at long wavelength with the 
 IUE observations (Kinney et al. \cite{kin}) and reaches 
 a maximum flux of about $3. \times 10^{-14}$
 erg cm$^{-2}$ s$^{-1}$ A$^{-1}$ at $\sim$ 1100 \AA. 
 At wavelengths shorter than 1000 \AA~ the flux 
 decreases rapidly and below 960 \AA~ the signal, if any,
 becomes extremely weak (Figure 1); actually it drops to below zero
 between the Lyman-series airglow lines with the 
 uniform background correction of 1 count cm$^{-2}$ s$^{-1}$ applied by
 the standard calibration pipeline. The faint level 
 between the redshifted Lyman break of the galaxy 
 911.7 (1+0.0448) = 952.5 \AA~ and about 960 \AA~ is
 probably due to the accumulation of Lyman series
 absorption lines in the galaxy itself. These lines   
 are resolved from Ly$\beta$ to Ly7 
 (Ly$\gamma$ and Ly$\beta$ are not shown in Fig 1).

\begin{figure}
  \resizebox{\hsize}{!}{\includegraphics{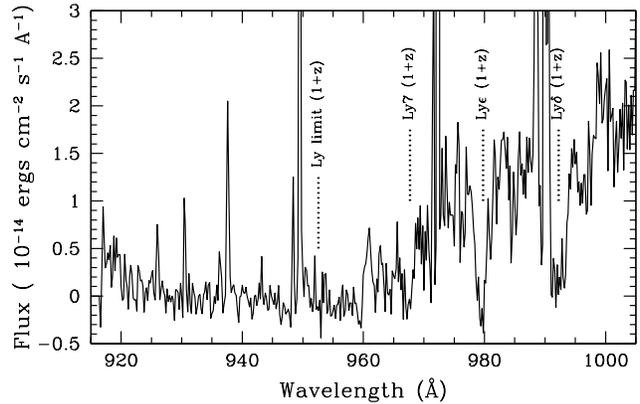}}
  \caption{{\it FUSE} spectrum of Mrk 54 resulting from the co-addition
  of 15 SiC2A spectra for a total exposure time of 27502 s and at 0.2 \AA~
  linear rebinning. The Lyman series absorption 
  lines resolved in Mrk 54 are marked except Ly6 confused with 
  the Ly$\gamma$ airglow line at 972.5 \AA.}  
\end{figure}

  For the purpose of discussing the 
  LyC radiation of the galaxy Mrk 54  
  we have concentrated  our analysis on the SiC channels 
  and specifically the SiC1B and SiC2A spectra.
  We have screened the raw ttag data of 
  segments 1B and 2A 
  for the presence of so-called bursts, using 
  for each of the 15 exposures both the images 
  of the events and the time series of the total number of counts.
  The latter has shown some increase of the count rate 
  at the end of a few exposures: 
  as these features were not
  typical burst events we decided not to remove them
  from the ttag data 
  (even for the most significant in exposure 13). We made no attempt
  at isolating night-only data.

\begin{figure}
\resizebox{\hsize}{!}{\includegraphics{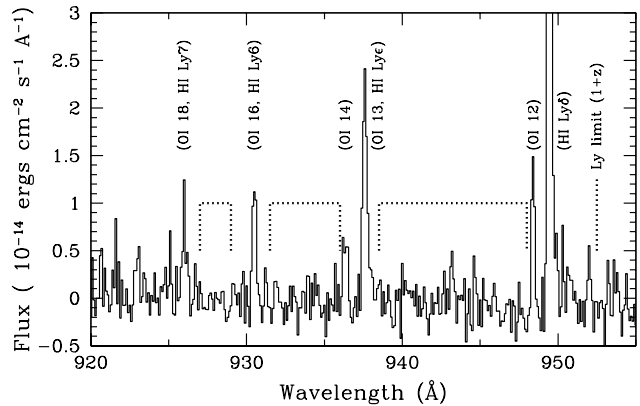}}
\caption{Details of the previous spectrum at 0.1 \AA~ rebinning,
   showing (dotted line) the three spectral windows (16 \AA~ total) used 
   for obtaining an upper limit to the LyC of Mrk 54. Tentative 
   identifications of the airglow lines based on Feldman et al. 
   (\cite{fel}) are indicated.}
\end{figure}

    In the absence of any clear signal from the redshifted        
  LyC  of the galaxy (shortward of 952.5 \AA) 
  we have tried to place an upper limit to this flux. A first step
  is to determine spectral windows     
  that are free from additional
  absorption by 
  neutral hydrogen in our Galaxy (Hurwitz et al. \cite{hur}). 
  As discussed by these latter authors, it is a complex 
  issue given the lack of information regarding
  low column density clouds at velocities departing 
  significantly from that of the bulk of the \ion{H}{i}.
  The examples given by Hurwitz et al. (\cite{hur}) and 
  by Lockman \& Savage (\cite{loc}) as well as the examination of
  the list of high-velocity clouds of Stark et al. (\cite{sta})
  lead us to conclude that removing $\pm$ 200 km s$^{-1}$
  on either side of the Lyman series absorption lines 
  is probably safe enough. This selection removes the 
  \ion{H}{i} nightglow lines sitting right at the 
  rest wavelengths of the potential absorption lines.
  A comparison of our spectrum with 
  the reference airglow spectrum of Feldman et al. (\cite{fel}) 
  shows that we 
  should also avoid a few \ion{O}{i} lines close to \ion{H}{i} lines and 
  discard a slightly larger wavelength domain 
  shortward of each Lyman series absorption lines. This 
  corresponds to a velocity of $\approx -$ 500 km s$^{-1}$,
  making our previous limit even safer on the side of negative velocities.  
  With the constraints adopted it is not possible  
  to extend our study shortward of Ly7 (926.2 \AA).
  We practically end up with
  a window of 16 \AA~ total, split into three domains,    
  927 -- 929 \AA, 931.5 -- 936 \AA~ and 938.5 -- 948 \AA~
  by the Lyman series lines Ly6  930.7 \AA, Ly$\epsilon$
   937.8 \AA~ and Ly$\delta$ 949.7 \AA~ (Figure 2).
  We have calculated the total raw counts in this window 
  for each of the 15 SiC1B and 15 SiC2A extracted spectra.

  These raw counts can be compared with those obtained 
  in reference zones, defined in the 15 images of segments
  1B and 2A built from the corresponding raw ttag data (Figure 3).
  These reference zones, 
  supposed to represent the background at the time 
  of observations, have been searched 
  as close as possible from the detector areas where the sky observations
  are made. Our search was guided by the examination of 
  profiles integrated over a large number 
  of lines or columns of the co-added (15) images 
  in order to avoid
  detector edge effects and  
  airglow lines in the smaller apertures. We end up with three 
  areas of respectively 542101, 1023701 and 737751 pixels in the 
  1B segment and 322361, 610151 and 443631 pixels in the 2A segment.   
  In these zones the counts are found to follow Poisson statistics 
  extremely well.

  \begin{figure}
  \resizebox{\hsize}{!}{\includegraphics{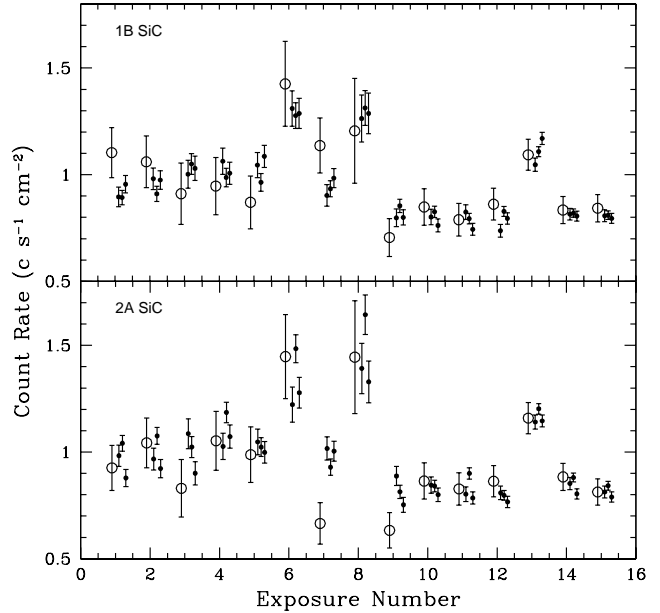}}
  \caption[] 
  {Comparison of
  the sky raw counts (open circle) in the 16 \AA~ window 
  defined in the text with 
  the background counts 
  in three reference zones (solid dots). These zones are 
  displayed in the same relative order for each 15 exposures. 
  For the purpose of comparison all counts have been expressed 
  into units of counts cm$^{-2}$ s$^{-1}$  
  using the respective exposure times, 
  the surface of the extracting windows for the sky counts, 
  the number of pixels and their linear 
  size (respectively 5.98 $\times$ 9.08 $\mu$m
  for the 1B segment and 5.96 $\times$ 14.75 $\mu$m for the 2A
  segment) for the background zones. 
  1$\sigma$ Poisson error bars are given.}
  \label{figstat} 
\end{figure}

  Three features stand out from Figure 3. First, given their 
  (1 $\sigma$) error bars the three background 
  measurements are neither significatively different in each exposures
  nor show systematic variation pattern. This is taken as a good 
  indication that the background is reasonably uniform 
  at least at the scale and in the zones we are using.  
  In contrast there are significant variations of the background
  measurements (departing from purely Poisson statistics) from exposure
  to exposure that can be understood by difference of observing conditions.
  Third, the target measurements follow the same trend as the background  
  from exposure to  
  exposure (except may be for exposure 7) and, given their larger 
  error bars, do not appear significantly different from the background
  measurements.

  These findings justify the use of 
  our evaluation of the background in each image 
  for correcting the sky measurement.  
  The error bars on the resulting net counts in each exposure
  combine quadratically the error bar on the sky measurements and those 
  on the background scaled to the surface of the sky measurements;    
  they are dominated by the uncertainties on the sky measurements based
  on a smaller number of pixels. The error bars of each 15 exposures 
  are then combined 
  quadratically and divided by the total exposure time to give the 
  dispersion on the net count rate.
  We obtain a mean net count rate of 
    1.32 $\times$ 10$^{-3}$ count s$^{-1}$ with a 1 $\sigma$ dispersion of 
    1.40  $\times$ 10$^{-3}$ count s$^{-1}$
and $-$1.13 $\times$ 10$^{-3}$ count s$^{-1}$ with a 1 $\sigma$ dispersion of 
    1.43 $\times$ 10$^{-3}$ count s$^{-1}$
  in the 16 \AA~ 
  window of segments 1B and 2A respectively. The fact that the 
  mean count rates are smaller than the dispersion confirms quantitatively the 
  trends discussed above. In these conditions the net count rate dispersion
  can be used to set an upper limit on the flux of Mrk 54
  at $\approx$ 900\AA~ (rest-frame).
  Combining the two segments and converting the counts into flux 
  (using an effective area of the order of 17 cm$^2$ 
   at $\sim$ 940 \AA~ according 
   to on-orbit performance reported by Sahnow et al. \cite{sah}) we get a  
  3 $\sigma$ upper limit of  4.7 $\times$ 10$^{-16}$ erg cm$^{-2}$ s$^{-1}$
   A$^{-1}$.  
  We have carried out all these calculations in counts rather than calibrated
  flux units in order to stay closer to basic count statistics.
  This advantage offsets the slight inaccuracy of
  converting counts into fluxes for the entire spectral window. 
   
\section{Analysis and discussion} 
 
\subsection{Correction to the upper limit}

  We need first to correct our upper limit flux   
  from any absorption that would
  not take place in the object but along its line of sight 
  in the intergalactic medium or in our Galaxy.
  At the low redshift of Mrk 54 the intergalactic medium opacity can be 
  neglected. As to the gas-phase absorption in our Galaxy, we have seen that
  the choice of our 16 \AA~ spectral window makes significant 
  absorption by neutral hydrogen unlikely.
  Most of the known interstellar metal lines,
  especially \ion{O}{i}, are also found to be avoided and the equivalent
  width of the corresponding absorption is negligible with respect to the 
  width of the window. For the molecular hydrogen 
  that {\it FUSE} has shown to be present along most of the extragalactic 
  lines of sight (Shull et al. \cite{shu2}) we have searched for the 
  presence of the most significant R(0) and R(1) lines 
  of the low-rotational levels of the 
  Lyman and Werner bands, following the identifications
  in the line of sight of ESO 141-055 (Shull et al. \cite{shu2}) 
  and the tables of
  Barnstedt et al. (\cite{bar}). 
  We found no clear identifications  even in
  the brightest parts of the spectrum. By comparison with the examples of
  ESO 141-055 and Mrk 876 (Shull et al. \cite{shu2}) we conclude that  
  $N(\mathrm{H}_2) <$ 1 $\times$ 10$^{18}$ cm$^{-2}$ in 
  the direction of Mrk 54.
  A difficulty of this comparison comes from  the fact that the spectra of 
  ESO 141-055 and Mrk 876 have better signal to noise ratio than Mrk 54
  (the objects are approximately two times brighter and 
  the exposure times two times longer). 
  We have used the on-line spectral simulator
  to explore how the detection is
  affected by these differences and found 
  that a H$_2$ column density of          
  1 $\times$ 10$^{18}$ cm$^{-2}$ would still 
  have been  detected in the conditions of 
  our Mrk 54 spectrum. If we give up any comparison 
  with observed spectra and use the simulator alone a lower limit 
  of 1 $\times$ 10$^{17}$ cm$^{-2}$ is found.
  We have then used the simulator to directly calculate the fraction of energy 
  that such H$_2$  column densities would absorb into our specific window of
  16 \AA; we found a fraction of 0.015 for 1 $\times$ 10$^{17}$ cm$^{-2}$  
  and 0.043 for the more conservative limit of 1 $\times$ 10$^{18}$ cm$^{-2}$.
  We have therefore applied no correction for H$_2$ absorption
  to our upper limit on the 
  LyC of Mrk 54. This case may well be a fortunate 
  circumstance given the galactic H$_2$ column densities 
  currently reported for extragalactic lines of sight 
  (Shull et al. \cite{shu2}, Vidal-Madjar et al.
   \cite{vid}).
  In the absence of a control of  
  the amount of  H$_2$ along the line of sight, that {\it FUSE} now offers,
  previous measurements may well have been affected by this problem.
  If accounted for, the H$_2$ contamination would have led
  to even less restrictive limits on the LyC
  escape fraction than established by Hurwitz et al. (\cite{hur}).

   Last, the Galactic dust extinction is a more severe and more uncertain 
   factor. In contrast to the current thinking for the past 17 years,
   it is very likely that a residual extinction 
   be present at high Galactic latitudes, even in the directions 
   with the lowest \ion{H}{i} column densities 
  (Schlegel et al. \cite{sch}). For 
   Mrk 54 we have adopted $E(B-V) = 0.015$ as given by the NASA Extragalactic
   Database on the basis of the latter reference. This color excess
   is consistent with a low fraction of hydrogen in the molecular 
   state (Savage et al. \cite{sav}) and therefore with previous 
   limits placed on the H$_2$ column density along 
   the line of sight to Mrk 54.  
   This results in a    
   foreground extinction of 0.29 mag in our far-UV window 
   (centered at $\sim$ 940 \AA) using an 
   extrapolation of the parameterized extinction law  of 
   Cardelli et al. (\cite{car}) shortward of 1000 \AA.
   Such an extrapolation is supported by the measurements of
   Buss et al. (\cite{bus}).
   We have therefore to account for a factor 1.306 of absorption 
   that does not occur in Mrk 54;
   our upper limit on the $f(900)$ flux of Mrk 54  is
   increased to 6.15 $\times$ 10$^{-16}$ erg cm$^{-2}$ s$^{-1}$
   A$^{-1}$.  
    
\subsection{Evaluation of the LyC escape fraction}      

      We can now proceed with the calculation of an upper limit 
   on the LyC escape fraction $f_{\mathrm{esc}}$ defined as the 
   fraction of emitted 900 \AA~ photons that escapes the galaxy 
   without being absorbed by interstellar material. 
   We need first to 
   have a relation between the $f(900)$ flux and 
   the total number of LyC photons since it is this latter quantity 
   that can be physically constrained by  
   the observed H$\alpha$ flux.
   Leitherer et al. (\cite{lei1}) have shown that there is a narrow relation
   relatively independent of star formation histories and initial mass
   functions 
   between the luminosity $L(900)$ at about 900 \AA~ and 
   the total number $N_{\mathrm{Ly}}$ of LyC
   photons of a burst population:  
    $\log (N_{\mathrm{Ly}}/L(900)) = 13.28 \pm 0.16$
   (photons \AA~ erg$^{-1}$).

   Assuming 0.45 as the number of H$\alpha$ photons per recombination
   (case B recombination at 10$^{4}$K), we write 
   the number of LyC photons $N_{\mathrm{Ly}}$, corrected by the 
   fraction of those escaping photoelectric absorption
   ($f1$) or trapped by dust 
   before ionization ($f2$), 
          $$ N_{\mathrm{Ly}} (1-f1-f2) = 7.34 \times 10^{11} 4 \pi D^2 
           f(\mathrm{H}\alpha) 10^{0.4 A(\mathrm{H}\alpha)}  $$ 
   with $f(\mathrm{H}\alpha$) the observed  H$\alpha$ flux 
  (in erg cm$^{-2}$ s$^{-1}$),
   $A(\mathrm{H}\alpha$) 
   the extinction 
   at H$\alpha$ in the observed galaxy and $D$ the distance (in cm).
    The luminosity $L(900)$ is
  $$ L(900) f_\mathrm{esc} =  4 \pi D^2 f(900) $$
  The continuum flux $f(900)$ 
   is corrected for gas-phase absorption in the Galaxy 
   and foreground dust extinction. It should be noted that   
   $f_\mathrm{esc}$ may be different from $f1$ because of 
    photons escaping photoelectric absorption but absorbed by dust 
   and the frequency dependence of photoelectric absorption.
   Combination of above equations gives:
   $$ f_\mathrm{esc} = {26  f(900) 10^{-0.4 A(\mathrm{H}\alpha)} 
            (1-f1-f2) \over {f(\mathrm{H}\alpha)}} \eqno(1)$$ 
   As we are dealing with an upper limit calculation
   in the case of Mrk 54 and we expect $f_\mathrm{esc}$, $f1$, and 
   $f2$ to be small it makes sense to write
   $$  {f_\mathrm{esc}} \leq {26  f(900) 10^{-0.4 A(\mathrm{H}\alpha)} 
                       \over {f(\mathrm{H}\alpha)}} \eqno(2) $$

    An H$\alpha$ flux of 
   2.6 $\times$ 10$^{-13}$ erg cm$^{-2}$ s$^{-1}$ has been obtained for Mrk 54
   from spectrophotometric observations    
   (Boselli \cite{bos}). An extinction $A(\mathrm{H}\alpha) = 0 $ 
   (although a value of 0.35 was obtained from the Balmer decrement 
   measured along one slit position)
   has been adopted in order to account for the uncertainty on the 
   H$\alpha$ flux in the upper limit calculation of equation (2).
   We finally obtain 
           $f_\mathrm{esc} < 0.062$ for Mrk 54.

      This result
   adds another significant limit 
   to three other cases with limits of 3.2\%, 5.2\% and 11\% as obtained 
   with HUT and the re-analysis by Hurwitz et al. (\cite{hur}) (Mrk 66 with a 
   limit of only 57\% is left over from this comparison). 
   All the values obtained so far are fully compatible 
   with current estimates
   in the range 2\% - 10\% based either on theoretical  
   models (Dove \& Shull \cite{dov1}, 
   Dove et al. \cite{dov2}) or the implications of 
   H$\alpha$ observations of the Magellanic Stream 
  (Bland-Hawthorn \& Maloney \cite{bla}) or NGC 3067 
  (Tumlinson et al. \cite{tum}). 

\subsection{Neutral hydrogen}

      Neutral hydrogen is expected in galaxies 
  where active star formation is taking place
  and its impact on the escape of ionizing photons 
  is thought to depend heavily on topology. 
  In Mrk 54 the neutral hydrogen is directly seen along the line of sight as
  Lyman series absorption lines (down to Ly7) at the redshift of the 
  parent galaxy. These features are accompanied 
  by a number of metal absorption-lines 
  with different possible origins (interstellar gas, 
  stellar photospheres and stellar winds) 
  as currently found in starburst galaxies (e.g. Heckman et al.
  \cite{hec} and references therein). Most of the analyses 
  so far have concentrated 
  on the features (including Ly$\alpha$)   
  accessible with {\it IUE} or the spectrographs of the {\it HST}. 
  To the best of our knowledge, the 
  discussion of features blueward of Ly$\alpha$ have been limited  
  to the \ion{O}{vi} + Ly$\beta$ + \ion{C}{ii} profile
  in the four starbursts observed with HUT (Gonzalez-Delgado et al.
  \cite{gon1}, \cite{gon2}).

     Leaving a full discussion of absorption features in Mrk 54 to 
  another context,  we have only tried to interpret
  the \ion{H}{i} Lyman series absorption profiles. 
  Within the uncertainties of background subtraction, 
  the lines appear almost black at their centers (this is less clear 
  with the Ly$\beta$ and Ly$\gamma$ lines than with the three lines displayed
  in Figure 1) and 
  indicate a large covering fraction of neutral hydrogen.  
  The determination of the neutral hydrogen column density implied 
  by the absorption profiles is however difficult because of the 
  low signal-to-noise ratio and  
  the possible blending with galactic interstellar lines
  (not to speak of nightglow emission lines) resulting in very uncertain 
   continuum levels: the profile fitting allows  
  column densities of the order of 10$^{21}$  cm$^{-2}$ with 
  low $b$ Doppler parameter
  as well as  low column densities  with $b$ up to $\sim 300$
  km s$^{-1}$. Large velocity spreads have already 
  been reported in nearby and high redshift starburst galaxies
  (e.g. Gonzalez-Delgado et al. \cite{gon2},
   Pettini et al. \cite{pet}).
   
     This finally leads us to place the discussion in the context of
  the large-scale outflows that have been revealed in the 
  interstellar media of starburst galaxies
  (e.g. Heckman \cite{hec2} and references therein) 
  and play a role in the escape of Ly$\alpha$ emission
  (e.g. Kunth et al. \cite{kun}). As the  Ly$\alpha$ emission often appears 
  redshifted with respect to absorption features
  the simplest interpretation is that 
  the only Ly$\alpha$ photons that escape unabsorbed are those 
  backscattered from the far side of the outflow whereas    
  the approaching part is seen in absorption
  against the stellar continuum (Pettini et al. \cite{pet}). In 
  this picture the LyC photons cannot escape in the same way as the 
  backscattered Ly$\alpha$ photons but it is likely that the galactic winds
  generate 
  holes in the \ion{H}{i} distribution 
  through 
  which  LyC photons  
  can also escape in different directions. 
  A more complete model, 
  including time evolution effect and accounting for the 
  variety of Ly$\alpha$ absorption and emission profiles 
  observed by Kunth et al. (\cite{kun}), has been developed by 
  Tenorio-Tagle et al. (\cite{ten}); in the phase where the 
  conical \ion{H}{ii} region extends to the galaxy outer edge, 
  LyC and Ly$\alpha$ photons should escape in the same 
  direction. 
  With only upper limits as obtained so far, 
  and only five nearby galaxies,
  it is difficult to constrain one particular model 
  by examining how the LyC escape fraction correlates with 
  the Ly$\alpha$ escape and the blend of absorption and emission 
  at Ly$\alpha$; for instance the 
  two objects with the tightest LyC escape fraction upper limit    
  (Mrk 496 and IRAS 08339) have both Ly$\alpha$ in emission. The only 
  sure conclusion of this discussion is the role of anisotropies that
  increase the randomness of the LyC escape fraction for the 
  observer.

\section{Comparison with the Lyman break galaxies}

   The upper limits ($<$ 10\%) obtained at z $\approx$ 0 
   are in stark contrast with 
   a LyC escape fraction larger than 50\% reported by Steidel et al.
   (\cite{ste}) from the detection of the       
  LyC radiation in a composite spectrum of Lyman break galaxies
  at z = 3.4. In Steidel et al.  
  the  LyC escape fraction is normalized  
  by the fraction
  of 1500 \AA~ photons that escapes. The surprisingly large value ($>$ 50\%) 
  results from the fact that the $f(1500)/f(900)$ flux ratio 
  is, after the necessary correction for the intergalactic medium 
  opacity at high redshifts, close to that
  predicted from spectral synthesis models without any LyC 
  self-absorption from neutral hydrogen in the galaxy.   
   The escape 
   fractions used at low and high redshifts are therefore   
   different both in reference wavelength
   (the escape fraction for nearby galaxies refers to H$\alpha$ vs. 
    1500 \AA~ for the Lyman break galaxies) and in the 
    treatment of dust extinction
   (dust extinction is accounted for at H$\alpha$ in nearby galaxies 
   whereas it is not for the Lyman break galaxies). 
   Before trying to quantify this difference 
   between the escape fractions, we have first examined a 
   quantity that is directly observed in both cases,
   the flux ratio $f(1500)/f(900)$.

\subsection{The $f(1500)/f(900)$ flux ratio of nearby galaxies}

  We have calculated 
  the $f(1500)/f(900)$ flux ratios in Mrk 54 and 
  the four nearby star-forming 
  galaxies observed with HUT (Table 1). For the comparison with the high-z
  galaxies to be meaningful we have to use the $f(900)$ upper 
  limit fluxes  corrected for gas-phase absorption and foreground
  dust extinction. As for Mrk 54, the latter correction has been based    
  on the $E(B-V)$ of Schlegel et al. (\cite{sch}) given in the 
  NED. These corrections are slightly different 
  from those of Leitherer et al. (\cite{lei1}) but the conclusions
  are not changed. All steps of calculations are detailed in 
  the notes to Table 1.
  The $f(1500)/f(900)$ flux ratios of the nearby objects
  (Table 1 column 7) are found to be larger than the value obtained 
  by Steidel et al. (\cite{ste}) from their composite spectrum of
  Lyman-break galaxies 
  (note that their value of 4.6 translates into a value of 1.7 in the units
  of Table 1).  This is especially true for Mrk 54 
  essentially because no correction for \ion{H}{i} absorption 
  has to be applied.

\begin{table*}
   \caption[]{Calculation of the 
             $f(1500)/f(900)$ ratio in nearby galaxies (the ratio of 4.6
          reported by Steidel et al. is 1.7 in the units of the table)}
         \label{comparison}
         \begin{tabular}{lcccccc}
         \hline
%           & & & & & &   \\
         Object  &  $f(900)$ & $E(B-V)$ & $f_{c}(900)$ & $f(1500)$ &
                       $f_{c}(1500)$ & $f_{c}(1500)/f_{c}(900)$\\
                 &  (2) & (3) & (4) & (5) & (6) & (7) \\
         \hline
%          & & & & & &   \\
          IRAS08339+6517 & $< 1.7 \times 10^{-15}$ &  0.092 & 
          $< 8.8 \times 10^{-15}$ & $2.8 \times 10^{-14}$ &
           $5.6 \times 10^{-14}$  & $> 6.4$\\
           Mrk 1267 & $< 1.4 \times 10^{-15}$ & 0.034 &
          $< 2.6 \times 10^{-15}$ & $8.5 \times 10^{-15}$ &
           $1.1 \times 10^{-14}$  & $> 4.2$\\
           Mrk 66   & $< 1.6 \times 10^{-15}$ & 0.012 & 
          $< 2.0 \times 10^{-15}$ & $7.5 \times 10^{-15}$ &
           $8.2 \times 10^{-15}$  & $> 4.1$\\
           Mrk 496  & $< 1.6 \times 10^{-15}$ & 0.020 & 
          $< 2.3 \times 10^{-15}$ & $8.0 \times 10^{-15}$ & 
           $9.3 \times 10^{-15}$  & $> 4.0$\\
           Mrk 54   & $< 4.7 \times 10^{-16}$ & 0.015 & 
          $< 6.2 \times 10^{-16}$ & $2.2 \times 10^{-14}$ &
           $2.5 \times 10^{-14}$  & $> 40 $\\
         \hline
        \end{tabular}
        \begin{list} {} {}
       \item Notes: all fluxes are in erg cm$^{-2}$ s$^{-1}$ A$^{-1}$.
       \item Col(2) Corrected for gas-phase absorption (Table 1 , column 12 of 
      Hurwitz et al. \cite{hur} for the galaxies observed with HUT). 
          Col(3) Foreground extinction from the NASA Extragalactic Database. 
          Col(4) Corrected for foreground extinction 
          with $A = 19.36 E(B-V)$, resulting from 
          the extrapolation to $\approx$ 940 \AA~ of the
          extinction law of Cardelli et al. (\cite{car}).
          Col(5) From the HUT or IUE spectra (Leitherer et al. \cite{lei1},
          Kinney et al. \cite{kin}). 
          Col(6) Corrected for foreground extinction with $A = 8.26 E(B-V)$.
  
       \end{list} 
\end{table*}

  Although the LyC photons escaping is probably 
  highly variable from galaxy to 
  galaxy, the number of nearby objects is sufficient to suggest 
  that the escape is 
  easier at redshift $\approx$ 3.4 than at low redshift.
   This trend  
  is now supported by similarly large $f(1500)/f(900)$
   flux ratios recently reported  
  by Ferguson (\cite{fer}) for seven galaxies 
  at z $\approx$ 1 in the HDF. 
   The UV luminosity may also be a factor in the sense that higher luminosity 
  at $\approx$ 1500 \AA~ would imply more photons for ionizing the gas
  and possibly easier escape. 
  The nearby galaxies have UV luminosities ranging      
  from $7.2 \times 10^{28}$ erg s$^{-1}$ Hz$^{-1}$ at $\sim$ 1500 \AA~
  for Mrk 54 to ten times less. However this largest luminosity is 
  similar to 
  the demagnified luminosity  
  of cB58 (Pettini et al. \cite{pet}) reported 
  by Steidel et al. as representative of the galaxies chosen for 
  the composite spectrum.

\subsection{The $f(1500)/f(900)$ flux ratio of Ly break galaxies}

     We have also examined the $f(1500)/f(900)$ flux
  ratio obtained for the Lyman break galaxies
  by Steidel et al. after correction for the IGM opacity. This ratio
  implies an escape fraction (when normalized to 1500 \AA) $\geq$ 0.5.
  A smaller escape fraction in the absolute sense 
  might be obtained if only a 
  fraction of the 1500 \AA~ photons escape 
  without being absorbed by dust. This argument is given by 
  Steidel et al. and used by Loeb \& Barkana
  (\cite{loe}) to infer an escape fraction of $\sim$ 10\%, close to 
  the limits obtained for nearby galaxies.  
  There is however a serious limitation with this explanation: 
  a significant dust extinction at 1500 \AA~ would likely imply
  a larger dust extinction at 900 \AA~ and in turn a   
  smaller intrinsic $f(1500)/f(900)$ ratio,
  uncomfortably small with 
  respect to those given by stellar population synthesis models
  (e.g. Bruzual \& Charlot \cite{bru}, Charlot \cite{cha},  
  Leitherer et al. \cite{lei2}). 
  The flatness (in f$_{\nu}$) of the composite spectrum obtained by 
  Steidel et al. also suggests a low extinction according to the 
  correlation between the ultraviolet extinction and the 
  slope $\beta$ of the far-UV spectrum (Calzetti et al. \cite{cal}). 
  The result of Steidel et al. is puzzling in another respect. If 
  the escape of LyC photons is anisotropic as discussed above 
  and suggested by Pettini et al. (\cite{pet}), it would be difficult 
  to have only the cases with favorable directions 
  appear in a composite spectrum. It is also true that the galaxies 
  contributing to the composite spectrum have been selected   
  using criteria that would favor the least extinguished and perhaps
  lowest covering fraction.

\subsection{Quantifying the difference between escape fractions}

   The LyC escape fraction appears in equation (2) as the 
  $f(900)$ flux divided by the H$\alpha$ flux and is defined by Steidel
  et al. as the $f(900)$ flux normalized by the fraction of 1500 \AA~
  photons that escapes.  It is tempting to establish a relation 
  between these two forms of the escape fraction, 
  independently of the $f(900)$ flux. Equation (2) can 
  be written as (neglecting the factors $f1$ and $f2$)   
   $$   f_{\mathrm{esc}} = 26{f(900)\over f(1500)} 
    {f(1500)\over f(\mathrm{H}\alpha)}
    {10^{-0.4 A(\mathrm{H}\alpha)}} $$ 
  On the other hand the escape fraction as defined by Steidel et al. 
  is
  $$   {f_{\mathrm{esc}2}} \approx 1.1 {f(900) \over f(1500)}  $$
   with 1.1 being the      
  $f(1500)/f(900)$ ratio (in units of Table 1) as
  expected from stellar synthesis models and adopted
  by Steidel et al. From the ratio of the two equations we get
  $${f_\mathrm{esc2} \over  f_\mathrm{esc}} \approx {0.04} 
      {f(\mathrm{H}\alpha) \over f(1500)} 10^{0.4 A(\mathrm{H}\alpha)} $$
  The ratio of the two escape fractions appears as depending 
  on the H$\alpha$ to far-UV (1500 \AA) ratio, a quantity more 
  currently measured than the flux at 900 \AA.
  The latter ratio (expressed in \AA~ with an UV flux in erg cm$^{-2}$ 
  s$^{-1}$ A$^{-1}$)
  can take in principle 
  any value from near zero (a burst  
  older than 7 Myrs without any ionizing star) to very large 
  (dust extinction). However,  
  it is currently measured in the range 10 -- 40 \AA~
  (e.g. Buat et al. \cite{bua2}, Meurer et al. \cite{meu}, 
   Sullivan et al. \cite{sul}, Bell \& Kennicutt \cite{bel})  
  for the bright, low-extinction 
  and UV-selected star-forming galaxies 
  as we are dealing with in this
  paper. The calibrations of Kennicutt (\cite{ken}) 
  for the far-UV and H$\alpha$ star formation rates correspond to 
  an  H$\alpha$ to far-UV ratio of 13 \AA~ and the model 
   calculations of Glazebrook et al.
  (\cite{gla}) encompass the range 6 -- 17 \AA~
  (extinction not included). 
   The $f_{\mathrm{esc2}}/f_{\mathrm{esc}}$ ratio is therefore found in the 
   range 0.4 -- 5,   
   assuming an extinction at H$\alpha$ in the range 0 -- 1.2
   (Kennicutt \cite{ken}). 
   This ratio can be calculated for the nearby galaxies and indeed 
    falls into the expected range. The large range obtained shows 
   the difficulty of comparing escape fractions with different
   definitions.

\vspace{0.3cm}   

        In conclusion, the difference between the large 
   LyC escape fraction 
   reported at high redshift and 
   the current upper limits in nearby galaxies must be interpreted 
   with caution because of the difference of the definitions of  
   the escape fractions used.
   In contrast the    
   $f(1500)/f(900)$ flux ratios in nearby star-forming galaxies
     suggest   
   that the  LyC photons escape, at least in the direction of the 
   observer, less easily from 
   nearby galaxies than from high redshift galaxies.    
   More and better measurements of the Lyman continuum
   of nearby galaxies are needed  
   to distinguish a possible evolution effect from selection effects.
    
\section{Implications}

    The stellar contribution to the hydrogen-ionizing background  
  has been investigated by a number of authors and its intensity 
  anticipated as a function of the LyC escape fraction. 
  These results can therefore be briefly revisited in the light 
  of present data and especially the $f(1500)/f(900)$ ratio.  

  With calculations similar to those of Madau et al.
  (\cite{mad2}) and Madau (\cite{mad3}), Steidel et al. 
  directly use their measured $f(1500)/f(900)$ ratio 
  to translate the luminosity function of Lyman break galaxies (at 1500 \AA~)
  into a distribution of Lyman continuum luminosities. They find
  that the 900-\AA~ emissivity of galaxies at z $\sim$ 3
  exceeds the contribution of QSOs by a factor of 5. This result supports   
  the role of star formation in 
  the early re-ionisation of the hydrogen but relies on the assumption that 
  the measured $f(1500)/f(900)$  ratio 
  is characteristic of the whole population of Lyman-break galaxies.

  At low redshifts 
  Giallongo et al. (\cite{gia}) and Shull et al. (\cite{shu1}) 
  have shown that the galaxy contribution to the ionizing 
  background rivals that of QSOs for an escape fraction of 5\%. 
  Such a value is compatible with the present set of upper limits 
  obtained with nearby star-forming galaxies.
  In the same vein as above, an upper limit on 
  the 900-\AA~ emissivity of galaxies at z $\approx$ 0 can be 
  directly translated from the far-UV luminosity density of
  Sullivan et al. (\cite{sul}), using  
  the $f(1500)/f(900)$ ratios in Table 1. 
  As the data of Sullivan et al. (\cite{sul}) are at 2000 \AA~ and a 
  mean redshift of 0.15, we have first used relations 
  established by Cowie et al. (\cite{cow2})
  to get the emissivity $0.9 \times 10^{26}$ 
  erg s$^{-1}$ Hz$^{-1}$ Mpc$^{-3}$ at 1500 \AA~ and 
  z $\approx$ 0 (this would be $0.7  \times 10^{26}$ with
  an $(1+z)^4$ evolution factor). 
  With the tightest constraints of Mrk 54 in Table 1
  we get a 900-\AA~ luminosity density $<$ 
  $8 \times 10^{23}$   erg s$^{-1}$ Hz$^{-1}$ Mpc$^{-3}$, 
  which is close to the emissivity due 
  to galaxies evaluated 
  by Shull et al. (\cite{shu1}) for an escape fraction of 5\%
  that would provide a comparable background to AGN.

    All the above calculations raise the issue on how 
  the value used for the LyC escape fraction is representative of
  the whole population of galaxies. Given all the factors 
  (distribution of neutral gas, orientation) that make this 
  parameter random, 
  it appears a crude oversimplification to use 
  an unique value for all galaxies. At the same time,
  any effort to understand the
  variations of this parameter 
  should carry the same level of difficulty as 
  establishing the  LyC luminosity function itself.
  
  As to the present calculations on Mrk 54 and comparison with other 
  nearby galaxies, it is likely, from the mere fact that 
  the  LyC escape fraction is random, that any  
  average value should lie below the current upper limits. In addition,
  any dust extinction decreases 
  the LyC escape fraction below the value expected 
  from neutral hydrogen absorption alone (see eq 1 or 2). 
  The nearby galaxies 
  under study, selected with significant UV flux for the prospect 
  of detecting ionizing radiation,
  are probably not representative of the whole population in terms of 
  dust extinction. Such an effect would be present  
  when it comes to using an average  escape fraction 
  with an H$\alpha$
  luminosity density for 
  deriving the  
  LyC luminosity density of galaxies.

\section{Conclusions}
   
      The wavelength domain of the {\it Far Ultraviolet Spectroscopic 
  Explorer (FUSE)} down to 905 \AA~ allows us to observe the  LyC radiation 
  of low redshift galaxies above the limit of photoelectric 
  absorption by neutral hydrogen in our Galaxy. The high spectral 
  resolution of {\it FUSE} combined with an appropriate redshift selection 
  allows us to avoid 
  Lyman series absorption by neutral hydrogen in our Galaxy over 
  a reasonably large wavelength range. 
  The high spectral resolution of {\it FUSE} is also useful to identify 
  (and eventually to correct for) possible gas-phase absorption from
  molecular hydrogen and species other than hydrogen.  
  The observation of the star-forming 
  galaxy Mrk 54 (z = 0.0448) is reported here with the following results:

   (i) The LyC flux is not detected. An upper limit 
   can be reasonably evaluated by 
   comparison with the background in adjacent zones of the detector. 
   It is measured  
   over three spectral windows (16 \AA~ in total) that are free to 
   a high probability of any absorption by neutral hydrogen in our 
   Galaxy. No correction for molecular hydrogen absorption is needed
   according to the lack of the corresponding absorption features 
   in the spectrum at longer wavelengths. A correction for foreground
   dust extinction is performed.
  
   (ii) By comparison with the H$\alpha$ flux of Mrk54, the resulting 
   upper limit on the  LyC flux 
   translates into an upper limit LyC escape fraction  
   $f_\mathrm{esc} < 0.062$. This result is compatible with current models
   for the escape of  LyC radiation in galaxies and 
   adds another significant limit to three other cases 
   with limits of the order or less than 10\%.

   (iii) A low escape of LyC photons 
   is expected in Mrk 54 from the large 
   covering fraction of neutral hydrogen revealed  by the       
   Lyman series absorption lines down to Ly7. The randomness of 
   the LyC escape fraction as resulting from the geometry
   (anisotropy in escape direction, hole in the neutral gas
   distribution) and large-scale
   galactic outflows is emphasized. No relation    
   has yet been established between the LyC and the Ly$\alpha$ emission
   escape in starburst galaxies.    

   (iv) The contrast between the 
   detection of significant LyC flux in a composite spectrum 
   of Lyman break 
   galaxies (Steidel et al. \cite{ste}) at z $\approx$ 3.4 
  and the fact that only 
  upper limits have been obtained in 
   nearby star-forming galaxies  
   is investigated. 
   The $f(1500)/f(900)$ flux ratio is found 
   larger in nearby galaxies than 
   in the composite spectrum of Steidel et al. (\cite{ste}). The possible 
   trend of LyC photons escaping more easily at high redshifts than at low 
   redshifts is now supported by the STIS data at z $\approx$ 1 of Ferguson
   (\cite{fer}) but more and better observations of nearby galaxies 
   are needed to distinguish an evolution effect from selection 
   and random effects. 
    
   (v) In addition to an  evaluation  
   from the H$\alpha$ emissivity and the LyC escape fraction,
   the 900-\AA~ emissivity of galaxies and potentially the 
   ionizing background radiation can now be derived from 
   the measured UV luminosity density and (UV to 900 \AA~) ratio. 

\begin{acknowledgements}
   We thank the {\it FUSE} Team at JHU for the successful operation 
  of such a complex instrument.
  J-M D thanks G. Kriss for providing specific scripts for reducing 
  {\it FUSE} data with IRAF, and  A. Boselli and G. Gavazzi for 
  their optical spectroscopic observation of Mrk 54. 
  This research has made use of the NASA/IPAC Extragalactic Database (NED)
  which is operated by the Jet Propulsion
  Laboratory, California Institute of Technology, under contract with 
  the National Aeronautics and Space Administration. 
\end{acknowledgements}

\end{document}